\documentclass[aps,prb,twocolumn,showpacs,superscriptaddress]{revtex4-1}
\usepackage{graphicx}

\begin{document}

\title{The influence of electron-doping on the ground state of (Sr$_{1-x}$La$_{x}$)$_2$IrO$_{4}$}

\author{Xiang Chen}
\affiliation{Department of Physics, Boston College, Chestnut Hill, Massachusetts 02467, USA}
\affiliation{Department of Materials, University of California, Santa Barbara, California 93106, USA}
\author{Tom Hogan}
\affiliation{Department of Physics, Boston College, Chestnut Hill, Massachusetts 02467, USA}
\affiliation{Department of Materials, University of California, Santa Barbara, California 93106, USA}
\author{D. Walkup}
\affiliation{Department of Physics, Boston College, Chestnut Hill, Massachusetts 02467, USA}
\author{Wenwen Zhou}
\affiliation{Department of Physics, Boston College, Chestnut Hill, Massachusetts 02467, USA}
\author{M. Pokharel}
\affiliation{Department of Physics, Boston College, Chestnut Hill, Massachusetts 02467, USA}
\author{Mengliang Yao}
\affiliation{Department of Physics, Boston College, Chestnut Hill, Massachusetts 02467, USA}
\author{Wei Tian}
\affiliation{Quantum Condensed Matter Division, Oak Ridge National Laboratory, Oak Ridge, Tennessee 37831, USA}
\author{Thomas Z. Ward}
\affiliation{Materials Science and Technology Division, Oak Ridge National Laboratory, Oak Ridge, Tennessee 37831, USA}
\author{Y. Zhao}
\affiliation{NIST Center for Neutron Research, National Institute of Standards and Technology, Gaithersburg, MD 20899, USA}
\affiliation{Department of Materials Science and Engineering, University of Maryland, College Park, Maryland 20742, USA}
\author{D. Parshall}
\affiliation{NIST Center for Neutron Research, National Institute of Standards and Technology, Gaithersburg, MD 20899, USA}
\author{C. Opeil}
\affiliation{Department of Physics, Boston College, Chestnut Hill, Massachusetts 02467, USA}
\author{J. W. Lynn}
\affiliation{NIST Center for Neutron Research, National Institute of Standards and Technology, Gaithersburg, MD 20899, USA}
\author{Vidya Madhavan}
\affiliation{Department of Physics and Frederick Seitz Materials Research Laboratory,  University of Illinois Urbana-Champaign, Urbana, Illinois  61801, USA}
\author{Stephen D. Wilson}
\email{stephendwilson@engineering.ucsb.edu}
\affiliation{Department of Materials, University of California, Santa Barbara, California 93106, USA}

\begin{abstract}
The evolution of the electronic properties of electron-doped (Sr$_{1-x}$La$_x$)$_2$IrO$_4$ is experimentally explored as the doping limit of La is approached.  As electrons are introduced, the electronic ground state transitions from a spin-orbit Mott phase into an electronically phase separated state, where long-range magnetic order vanishes beyond $x = 0.02$ and charge transport remains percolative up to the limit of La substitution ($x\approx0.06$).  In particular, the electronic ground state remains inhomogeneous even beyond the collapse of the parent state's long-range antiferromagnetic order, while persistent short-range magnetism survives up to the highest La-substitution levels.  Furthermore, as electrons are doped into Sr$_2$IrO$_4$, we observe the appearance of a low temperature magnetic glass-like state intermediate to the complete suppression of antiferromagnetic order. Universalities and differences in the electron-doped phase diagrams of single layer and bilayer Ruddlesden-Popper strontium iridates are discussed.  

\end{abstract}


\maketitle

\section{Introduction}
Resolving the electronic response to carrier substitution remains central to the study of the new class of materials manifesting a spin-orbit Mott (SOM) state---a Mott state stabilized by the cooperative interplay of strong spin-orbit coupling and on-site Coulomb repulsion \cite{PhysRevLett.101.076402,PhysRevLett.101.226402}.  Specifically, as the metallic regime is approached and the SOM state is destabilized, the continued relevance of correlation effects and their propensity to drive additional modes of symmetry breaking/electronic phase separation in SOM states remain poorly understood.  Of equal importance, understanding the microscopic doping mechanisms of SOM systems remains critical to resolving the underlying origins of the increasing number of novel properties reported in carrier-tuned SOM states.  

Of the families of SOM states recently uncovered, layered quasi-two-dimensional Sr$_2$IrO$_4$ (Sr-214) has received substantial attention\cite{PhysRevLett.101.076402}---partially due to theoretical predictions that this material may manifest an unconventional superconducting phase when doped\cite{PhysRevLett.106.136402, PhysRevLett.110.027002}.  In particular, model Hamiltonians of electron-doping in this Sr-214 compound have been mapped to those of hole-doping the Mott states of high temperature superconducting copper oxides, and empirical similarities between the 2D Heisenberg $S=\frac{1}{2}$ and $J_{eff}=\frac{1}{2}$ magnetic ground states of La$_2$CuO$_4$ \cite{PhysRevB.59.13788} and Sr-214 \cite{PhysRevLett.108.247212} have spurred further interest.  Experimentally, however, the evolution of Sr-214's properties upon electron substitution are surprisingly complex and poorly understood. 
   
Undoped, the SOM parent state of Sr$_2$IrO$_4$ manifests a robust charge gap of $E_{G}\approx 600$ meV \cite{PhysRevB.90.041102, PhysRevB.89.085125} and an antiferromagnetically ordered ground state \cite{Kim06032009}.  Charge transport data within this state shows a localized, tunneling form of conduction, which correlates to three-dimensional variable range hopping (VRH) at low temperatures \cite{0953-8984-18-35-008, PhysRevB.57.R11039}.  The evolution of the charge gap and transport behavior upon cooling is seemingly driven by a complex interplay between the underlying Mott gap and the onset of long-range antiferromagnetic (AF) order.  One manifestation of this interplay appears as an enhancement of the charge gap as the system is cooled through $T_{AF}$\cite{PhysRevB.86.035128, Mandrus}, suggesting an underlying ground state with mixed Slater-like and Mott character \cite{PhysRevB.89.165115, PhysRevLett.108.086403}.

A canted antiferromagnetic (CAF) state is stabilized in the parent state of Sr-214 below $T_{AF}\approx230$ K \cite{Kim06032009,PhysRevLett.110.117207}.  The onset temperature is limited by the weak interplane coupling between IrO$_2$ layers, while the strong in-plane coupling $J_{n.n.}\approx 60$ meV \cite{PhysRevLett.108.177003} drives short-range, quasi-two-dimensional correlations to persist to much higher temperatures \cite{PhysRevLett.108.247212}.  Once ordered, AF moments of $\mu_{AF}\approx 0.35\mu_B$ \cite{PhysRevB.87.144405} remain locked to the in-plane octahedral tilts \cite{0953-8984-25-42-422202}, generating a weak ferromagnetic response under an applied in-plane field and an eventual metamagnetic transition beyond a critical $\mu_0 H_{C}\approx 0.2$ T \cite{PhysRevB.57.R11039, Kim06032009}.  The conventional picture of this canted state is that, due to the near negligible easy-axis anisotropy within the $ab$-plane \cite{PhysRevB.80.075112}, the net moment due to canting can be rotated and rapidly polarized by a very weak in-plane field.  This complicates detailed analysis of static spin susceptibility data, and scattering data have provided the clearest picture of the continuous evolution of the AF order parameter below $T_{AF}$ \cite{Kim06032009,PhysRevB.87.144405,PhysRevB.87.140406}.       

An additional energy scale, not readily captured within scattering derived magnetic order parameters, has been observed in the low temperature magnetization and magnetotransport properties of both this single layer Ruddlesden-Popper (R.P.) iridate \cite{PhysRevB.84.100402} as well as its bilayer counterpart, Sr$_3$Ir$_2$O$_7$ (Sr-327) \cite{PhysRevB.86.100401,PhysRevB.66.214412}.  The extent to which these anomalies in magnetization and magnetotransport constitute signatures of a unique order parameter, however, remains an area of open investigation.  Recent muon spin relaxation measurements have for instance suggested this behavior's origin as the formation of oxygen moments driven by hybridization between Ir $5d$ and O $2p$ orbitals \cite{PhysRevB.91.155113}, and theoretical work has suggested the possibility of novel, mixed character, density wave formation in this compound \cite{2014arXiv1407.3440D}.           

Early experimental studies on single crystals have shown that electron doping Sr-214 results in a rapid collapse of the SOM state under only several percent of electrons doped per Ir-site \cite{PhysRevB.84.100402}.  Specifically, electron-doping via La-substitution was reported to stabilize a low temperature metallic-state at $4\%$ La-substitution, and the creation of oxygen vacancies was also reported to form a low temperature metallic state \cite{PhysRevB.82.115117}---although chemical inhomogeneity inherent to limited oxygen diffusion precludes an in depth study of the latter.  The means through which this metal-insulator (MIT) occurs and the evolution of electronic degrees of freedom as the SOM state is destroyed remain important open questions and are the focus of the present work. 

In this paper, we present a combined magnetotransport, bulk magnetization, neutron diffraction, and scanning tunneling microscopy (STM) study of the electronic phase behavior in electron-doped (Sr$_{1-x}$La$_x$)$_2$IrO$_4$ for $0\leq x\leq 0.06$.  Our measurements reveal the rapid suppression of the insulating state upon La-subtitution into Sr-214; however, in contrast to an earlier report \cite{PhysRevB.84.100402}, we observe the persistence of a weakly insulating state up to the doping limit of La in the lattice.  STM data collected on samples near this La-doping limit show the nanoscale coexistence of fully gapped and gapless regions and demonstrate that the MIT apparent from charge transport data is percolative in nature.  Neutron scattering data reveal that long-range AF order vanishes at $x\approx0.03$ while magnetotransport and static spin susceptibility data demonstrate that short-range correlations persist up to the highest doping level measured ($x=0.06$).  As the SOM phase is suppressed, low temperature anomalies in static spin susceptibility and magnetotransport data also identify an additional doping-dependent phase transition, $T_{F}$, between $13$ K and $9$ K consistent with the formation of a spin glass-like state.  The persistence of short-range AF order and electronic phase segregation as well as the appearance of a low temperature spin-glass state within La-doped Sr-214 provide a notable contrast to the global collapse of the SOM state in electron-doped Sr-327 \cite{hogan}.  

\begin{figure}
\includegraphics[scale=.375]{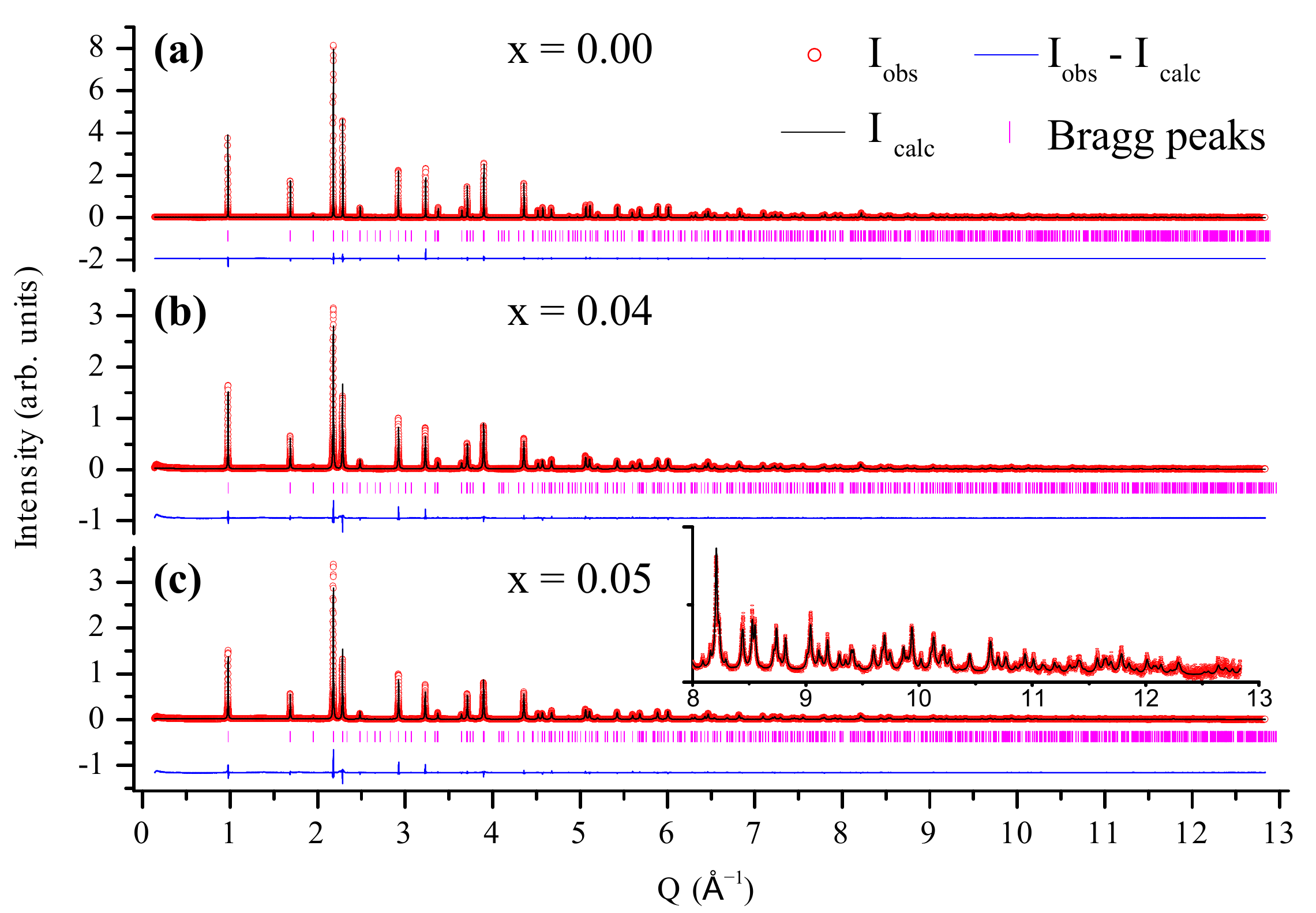}
\caption{Synchrotron x-ray powder diffraction data collected at 300 K with $\lambda=0.4136\AA$ for (a) polycrystalline x=0, (b) crushed single crystals with x=0.04, and (c) crushed single crystals with x=0.05.  Data were indexed and refined to the $I4_1/acd$ space group.  Inset in panel (c) shows magnified data and fit for x=0.05 at high Q.}
\end{figure}

\begin{figure}
\includegraphics[scale=.35]{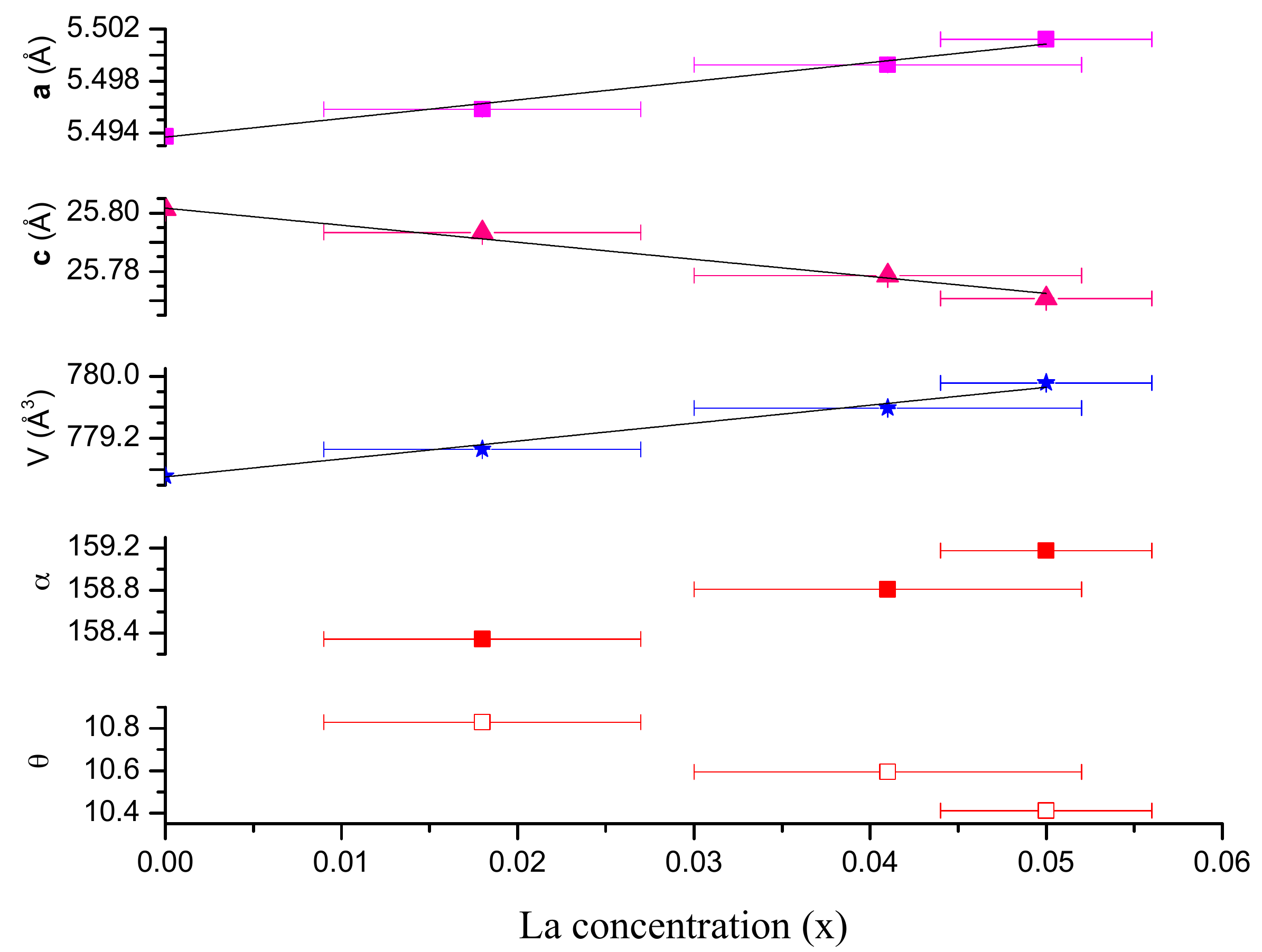}
\caption{Lattice parameters $a$ and $c$, unit cell volumes $V$, in-plane Ir-O-Ir bond angles ($\alpha$), and the resulting octahedral canting angles ($\theta$) refined for La-doped Sr-214 crystals. Solid lines are linear fits to the data.}
\end{figure}

\section{Experimental Details}
Samples were grown via previously established crucible-based flux methods \cite{Kim06032009, PhysRevB.87.144405}.  Single crystals of (Sr$_{1-x}$La$_x$)$_2$IrO$_4$ were grown in platinum (Pt) crucibles using SrCO$_3$ (99.99$\%$, Alfa Aesar), La$_2$O$_3$ (99.99$\%$, Alfa Aesar), IrO$_2$ (99.99$\%$, Alfa Aesar), and anhydrous SrCl$_2$ (99.5$\%$, Alfa Aesar) starting materials in a 2:1:6 molar ratio of (SrCO$_3$/La$_2$O$_3$):(IrO$_2$):(SrCl$_2$). Precursor powders were sealed inside Pt crucibles with Pt lids and placed inside outer alumina crucibles. Mixtures were heated to 1380 $^{\circ}$C, soaked for 10 hours at 1380 $^{\circ}$C, slowly cooled to 850 $^{\circ}$C at 4.5 $^{\circ}$C per hour, and then furnace-cooled to room temperature. Black plate-like single crystals with typical dimensions $1.5$ mm$\times 0.7$ mm$\times 0.2$ mm were then removed from the crucible with deionized water.

Representative crystals within each batch were chosen and ground into powder for initial measurement via powder x-ray diffraction in a Bruker-D2 Phaser diffractometer, and select crystals were later measured at the 11-BM beam line at the Advanced Photon Source at Argonne National Lab. No additional chemical phases or intergrowths were found within the resolution of these measurements.  Stoichiometry and the resulting La-concentrations were further checked via energy dispersive spectroscopy measurements, which defined the sample labeling used throughout the manuscript. Standard deviations of the La concentrations found within one growth batch and within one crystal were less than 1$\%$.

Transport measurements were performed both within a Quantum Design Physical Property Measurement System (PPMS) as well as using a resistivity platform within a Janis closed cycle refrigerator with a Lakeshore 370 AC Resistance Bridge.  Magnetization measurements were performed in a Quantum Design MPMS3 magnetometer.  Scanning tunneling microscopy (STM) measurements were performed in a helium bath cryostat with a custom built measurement platform.  All STM data was taken at 6 K.  Samples were cleaved at liquid nitrogen temperature and quickly inserted into the STM head at 6 K. $\frac{\delta I}{\delta V}$ spectra were acquired by using a lock-in technique while using electrochemically etched tungsten tips annealed in UHV and prepared on Cu(111).  

Neutron scattering measurements were performed on the HB-1A triple-axis spectrometer at the High Flux Isotope Reactor (HFIR) at Oak Ridge National Laboratory and on the BT7 spectrometer at the NIST Center for Neutron Research \cite{BT7}.  For experiments on HB-1A, a double bounce pyrolitic graphite (PG) monochromator and PG analyzer were utilized with a fixed $E_I=14.7$ meV, two PG filters before the sample, and collimations of $40^\prime-40^\prime-240^\prime-270^\prime$ before the monochromator, sample, analyzer, and detector respectively. For experiments on BT7, the beam collimations were $open-80^\prime-80^\prime-120^\prime$, and two PG filters were placed after the sample with a fixed $E_F=14.7$ meV. Crystals in these experiments were oriented within the [H0L] scattering plane and mounted within closed cycle refrigerators.  Uncertainties where indicated represent one standard deviation.

\section{Structural Characterization}
Select (Sr$_{1-x}$La$_x$)$_2$IrO$_4$ crystals with $x = 0.04$, $x = 0.05$, and polycrystalline $x = 0$ were crushed and characterized via x-ray diffraction.  Recent studies have revealed a subtle structural symmetry breaking \cite{PhysRevB.87.144405, PhysRevB.87.140406} which results in two unique Ir-sites per plane and a crystallographic space group of $I4_1/a$ \cite{PhysRevLett.114.096404}.  This oxygen octahedral distortion is most readily observed via single crystal neutron measurements or optical techniques, and the associated superlattice is not resolved in our synchrotron powder data.  As an approximation, the powder data were fit within the $I4_1/acd$ space group \cite{PhysRevB.49.9198} with the resulting structural refinements plotted in Fig. 1.  Crushed crystals near the La solubility limit contained only the single $I4_1/acd$ phase, confirming the absence of intergrowths of higher order members of the R.P. series.  

\begin{figure}
\includegraphics[scale=.4]{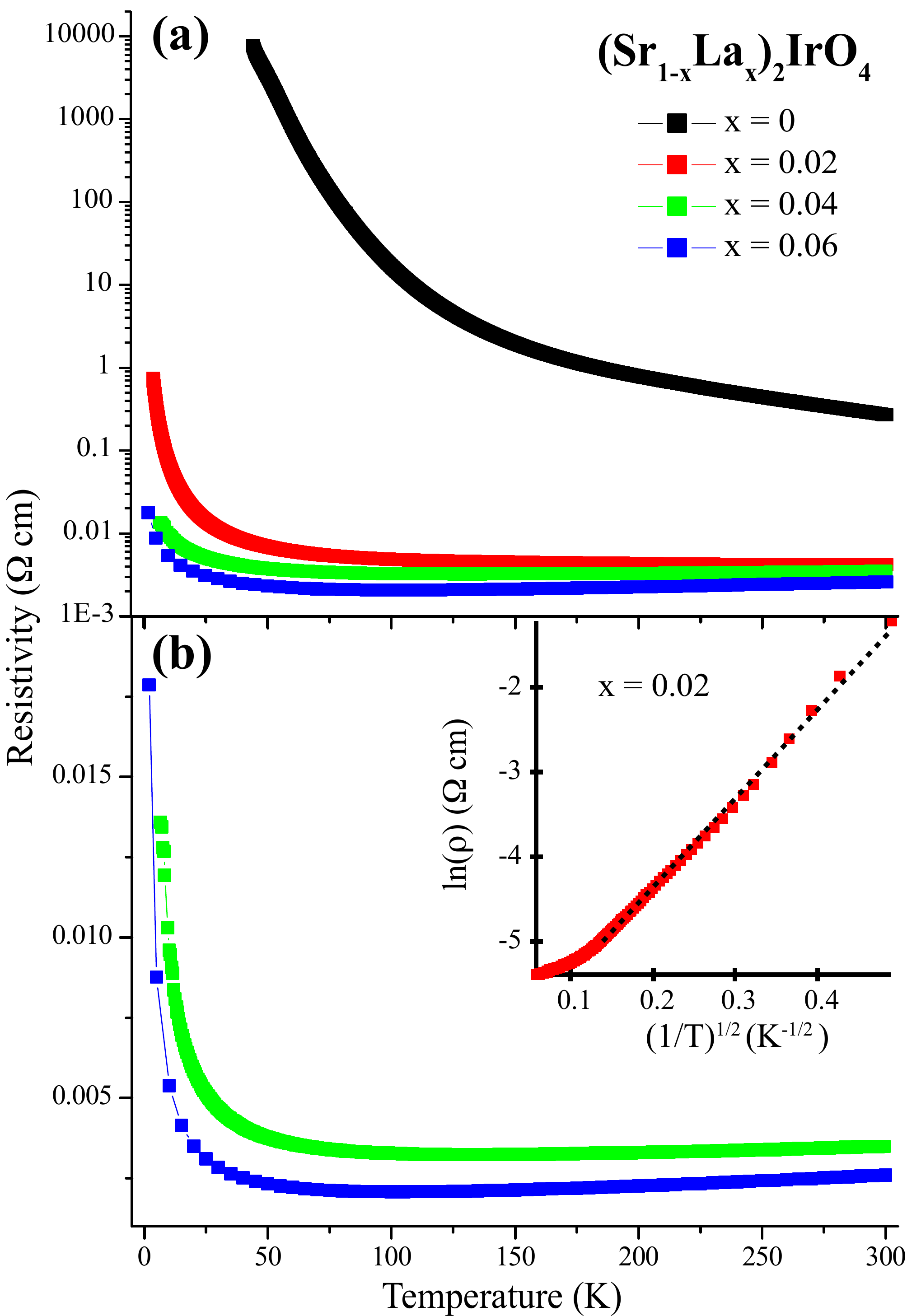}
\caption{(a) Temperature dependence of in-plane resistivity $\rho(T)$ for samples with $x=0, 0.02, 0.04$, and $0.06$. (b) x=0.04 and x=0.06 $\rho(T)$ data plotted on a linear scale.  Inset shows $ln(\rho)$ for $x=0.02$ plotted as a function of $T^{-1/2}$.  Dashed line is a linear fit as discussed in the text.}
\end{figure}

Lattice parameters, unit cell volumes, and bond angles of crushed La-doped Sr-214 crystals are plotted in Fig. 2.  Consistent with earlier reports, the Ir-O-Ir in-plane bond angle $\alpha$ slightly increases with La-substitution and the corresponding canting angle $\theta$ of the in-plane oxygen octahedra (ie. $\frac{180-\alpha}{2}$) is reduced \cite{PhysRevB.84.100402}.  The addition of donors to the system induces a swelling of the lattice volume, driven by the dominant expansion of the in-plane lattice constants accompanied by a reduction in the c-axis lattice parameter.  This primarily electronically-driven lattice response is comparable to that observed in (Sr$_{1-x}$La$_{x}$)$_3$Ir$_2$O$_7$ \cite{hogan} where an interplay between correlation effects and modified lattice deformation potentials dominate simple steric effects \cite{Janoti}.  However, unlike La-doped Sr-327, our neutron and x-ray scattering experiments on La-doped Sr-214 failed to detect any low temperature change in the lattice symmetry up the doping limit of $x=0.06$. 

\section{Charge Transport Measurements}
Charge transport data are plotted in Fig. 3 showing the evolution of the $ab$-plane resistivity as the La-content in Sr-214 is increased.  In substituting only a few percent of La into the system, resistivity data reveal a dramatic decrease in the room temperature resistivity by nearly two orders of magnitude.  Upon cooling, insulating behavior (defined here by $\frac{\partial\rho}{\partial T}<0$) persists in all samples at the lowest temperatures measured ($2$ K); however the magnitudes of the low temperature resistivity values approach those of disordered metals.  At the highest doping concentration measured, $x=0.06$, a temperature-driven switch from a high temperature metallic behavior to a low-temperature insulating upturn in resistivity was observed below $T\approx 100$ K (Fig. 3 (b)).  

The low temperature $\rho (T)$ of La-doped Sr-214 with $x=0.02$ is plotted in the inset of Fig. 3 (a).  Data from this lightly doped sample are best fit to a Efros-Shklovskii form of VRH \cite{0022-3719-8-4-003} with a characteristic temperature $T_{0}=112$ K---suggesting a qualitative departure from the large $T_{0}$ and three-dimensional Mott VRH behavior observed in the parent system \cite{0953-8984-18-35-008, PhysRevB.57.R11039}.  Alternatively, this functional form of transport behavior can be interpreted in the context of intergranular hopping in an electronically phase separated metal, similar to the case of doped cobalitites \cite{wu2005intergranular,sheng1973hopping}.  At higher doping levels, fits to this form break down as the MIT is approached and a high temperature metallic state appears.  The remnant low temperature upturn in $\rho(T)$ for $x\geq0.04$ may suggest weak localization effects, the opening of a low temperature pseudogap, or, alternatively, thermally induced shifts in percolation pathways within an inhomogeneous electronic ground state (discussed further in Sect. VIII).   

\section{Magnetization Measurements}
In order to explore the evolution of magnetic order in electron-doped Sr-214, bulk magnetization measurements were performed as La was progressively introduced into the Sr-214 matrix.  The canted AF spin structure of the parent material is known to generate a weak ferromagnetic response below $T=240$ K in the presence of an in-plane magnetic field \cite{PhysRevB.57.R11039}. $M(T)$ data collected under an $ab$-plane oriented field ($H_{ab}$) are plotted in Fig. 4 and show the evolution of this magnetic response as electrons are introduced into the system.  

In looking first at the parent compound, an initial ferromagnetic upturn below $T\approx240$ K is generated by biasing the nominally zero moment layered structure of canted planar moments.  Continued cooling causes this net magnetization to relax as the AF moment develops and the spin structure hardens.  This behavior is emblematic of a canted AF response; however the data also suggest an additional energy scale $T_{F}\approx26$ K, indicated by a cusp in the zero field cooled (ZFC) magnetization data.  

\begin{figure}
\includegraphics[scale=.35]{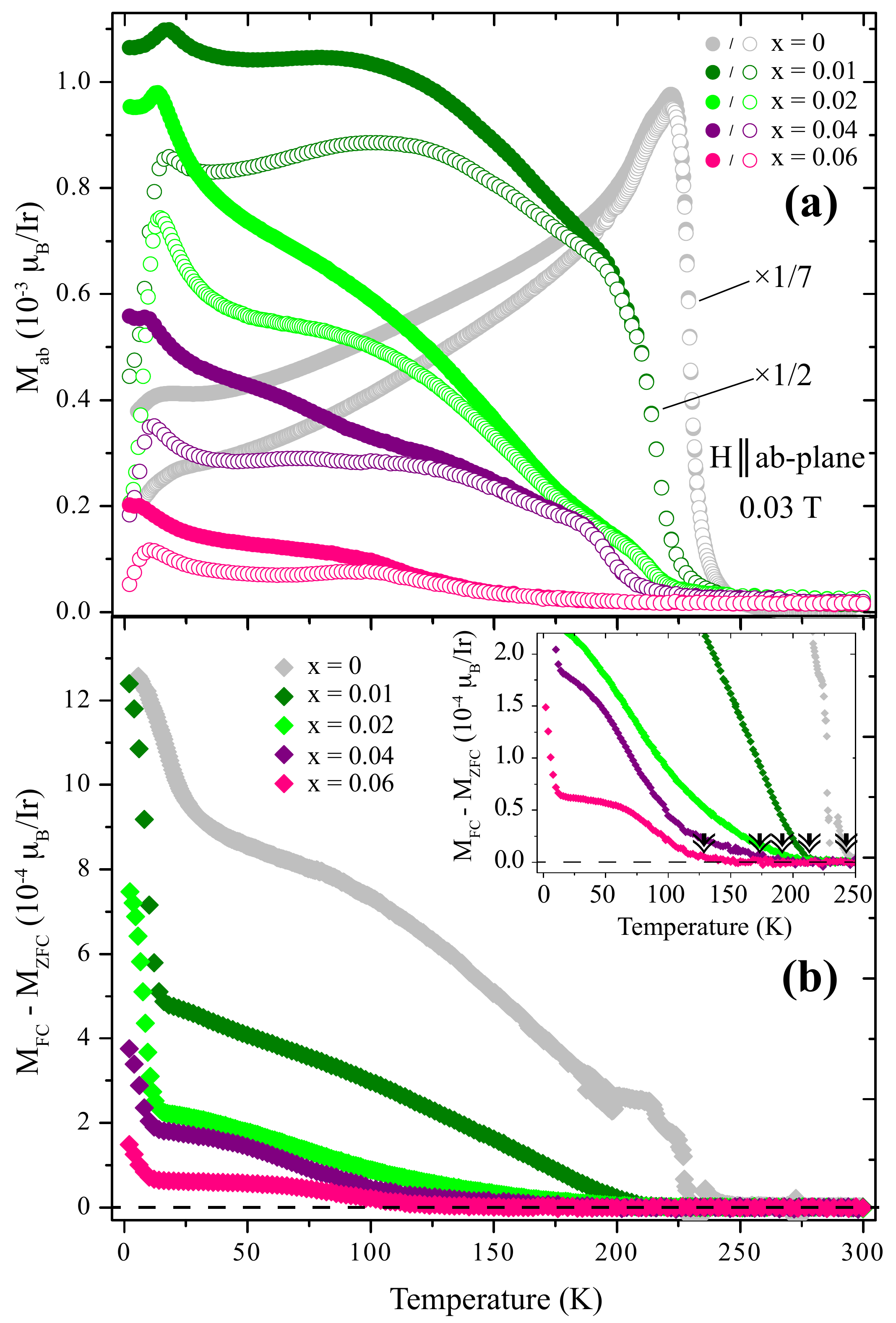}
\caption{Temperature dependent magnetization data collected under an in-plane applied field $\mu_0H_{ab}=0.03$ T. (a) Zero field cooled (open symbols) and field cooled (closed symbols) magnetization data showing the high temperature onset of static canted AF correlations at $T_{AF}$ and the low temperature onset of ZFC cusp at $T_{F}$.  For clarity, data for $x=0$ have been multiplied by $\frac{1}{7}$ and data for $x = 0.01$ have been multiplied by $\frac{1}{2}$ (b) Subtraction of FC and ZFC data in panel (a).  Inset shows the onset of irreversibility associated with canted AF ordering marked by double arrows.}
\end{figure}

As La is doped into the system, the static spin susceptibility is quickly suppressed; however signatures of magnetic order remain in all samples.  To better characterize the evolution of AF order, the irreversibility (ie. field cooled (FC) minus zero-field cooled (ZFC)) of the magnetization data presented in Fig. 4 (a) data are plotted in Fig. 4 (b).  From this data, it is apparent that the onset of irreversibility ($T_{AF}$) shifts to lower temperatures with electron doping and that the magnitude of the low temperature moment induced under fixed $H_{ab}$ is reduced. Data collected from the $x=0.06$ sample at the limit of La-substitution show a persistent weak ferromagnetic upturn and accompanying irreversibility, demonstrating the survival of static magnetic correlations at this limit. 

The low temperature cusp in ZFC data and inflection in irreversibility denoting $T_{F}$ also remains in all doped samples and shifts downward in temperature with increased electron concentration, from $T_F=13$ K for $x=0.01$ to $T_F=9$ K for $x=0.06$.  The magnitude of the irreversibility at $T_{F}$ is largest for the smallest doping $x=0.01$ and progressively diminishes with continued doping and the subsequent weakening of AF order.  To further explore this, $M(T)$ data from a representative $x=0.02$ sample were collected under various in-plane fields and the effective $\chi(T)=M/H$ plotted in Fig. 5.  Upon increasing $H_{ab}$, the irreversibility associated with the onset of AF order is suppressed as the critical field needed to align the canted moments of each layer is traversed ($\mu_0 H_c\approx > 0.2$ T) and the weak ferromagnetic moment is saturated. The cusp and irreversibility associated with $T_{F}$, however, persist as shown in Fig. 5 (b). The effective onset temperature of $T_{F}$ as seen via magnetization continues to decrease with increasing field prior its near complete suppression under $\mu_0 H\approx 5$ T.   

Low temperature M(H) data collected below $T_{F}$ for select La-doped samples are plotted in Fig. 6.  With increased electron concentration, the metamagnetic transition at $H_C$ inherent to the polarization of canted AF layers in Sr-214 is broadened and the saturated moment (approximated by $M(6T)$) rapidly diminishes.  An enhanced hysteresis at low La concentrations leads to a peak in the difference of hysteresis loops at finite fields as illustrated in Fig. 6 inset.  In metallic spin glass systems, the interplay between weak ferromagnetism and spin glass order is known to yield a similar response \cite{RevModPhys.58.801}, which when combined with the ZFC cusp observed at $T_{F}$ supports the notion of a spin glass state at low temperature in doped samples.  Meanwhile, the rapid suppression of the high field $M(H)$ with La-substitution reflects the doping-driven suppression of the parent AF magnetic state via a diminishing ordered moment or a rapidly reduced ordered volume fraction within a phase separated ground state \cite{dhital327}.  

\begin{figure}
\includegraphics[scale=.35]{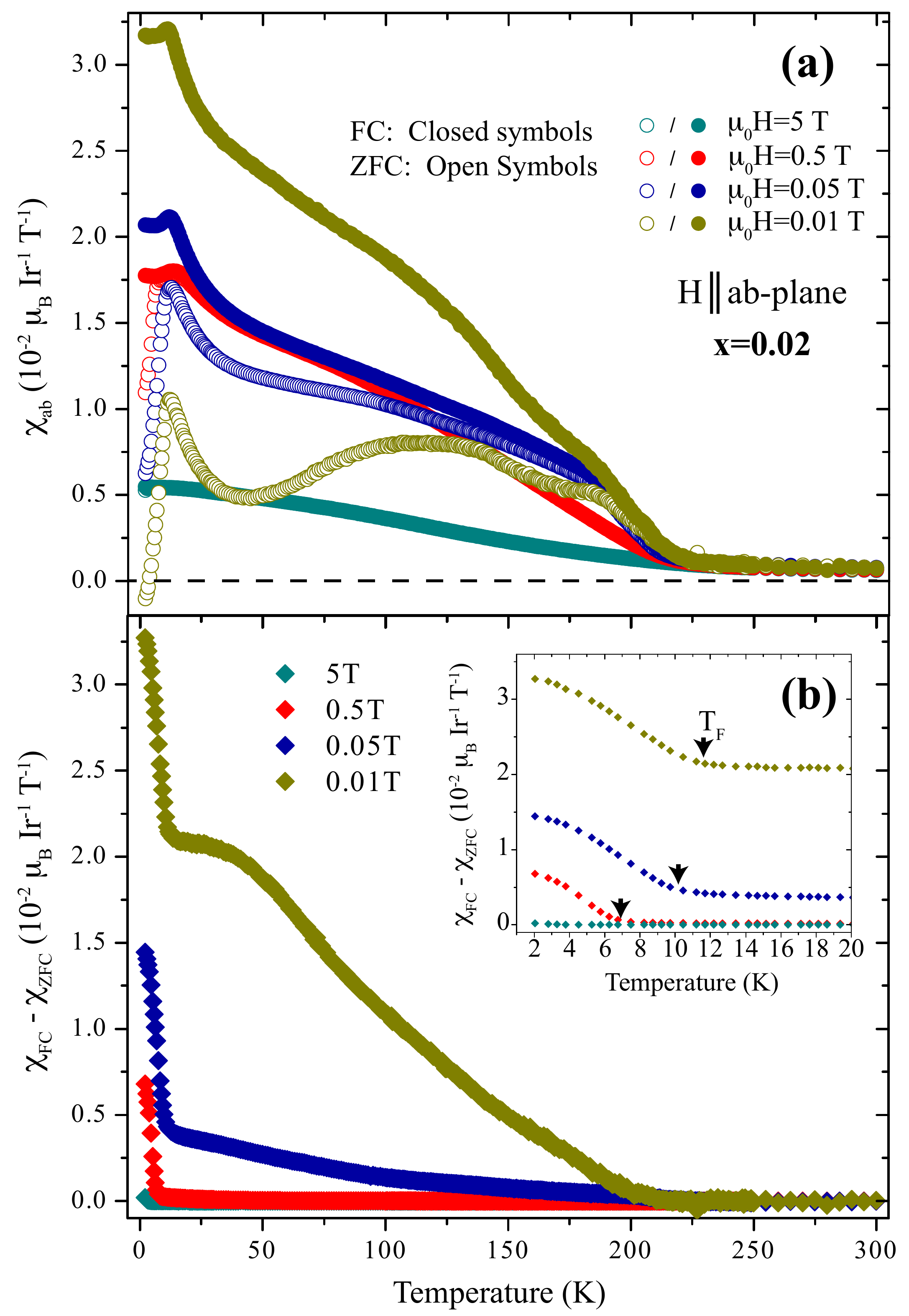}
\caption{Effective static spin susceptibility $\chi(T)=M(T)/H$ for the $x=0.02$ concentration under various $ab$-plane oriented fields. (a) $\chi(T)$ measured under both FC (closed symbols) and ZFC (open symbols) conditions. (b) The resulting irreversibility determined by the difference of FC and ZFC data plotted in panel (a).  Inset shows a zoomed region highlighting the field dependent irreversibility associated with $T_{F}$. The low temperature inflection associated with $T_{F}$ is denoted by a single arrow for each concentration.}
\end{figure}

\begin{figure}
\includegraphics[scale=.35]{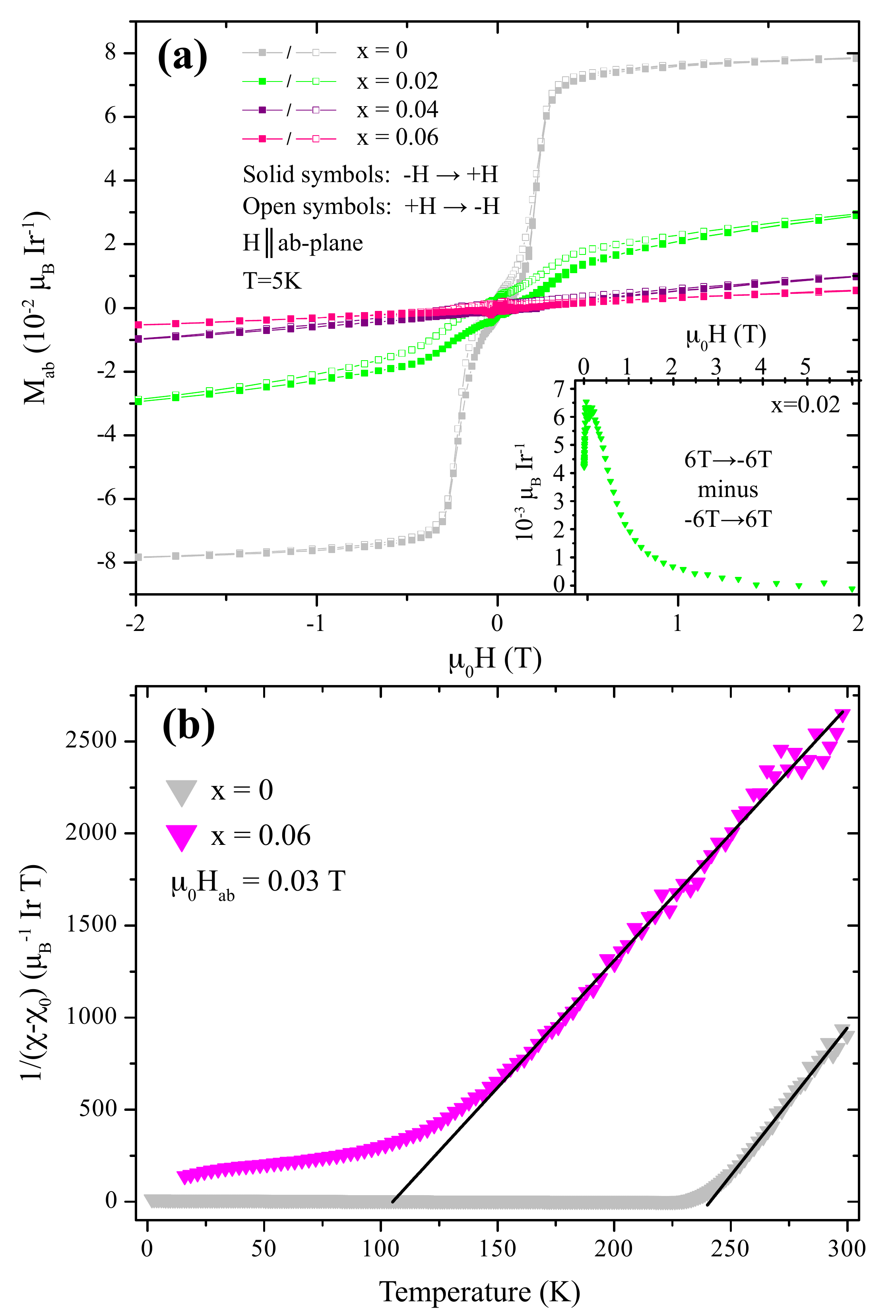}
\caption{(a) Magnetization as a function of field for La-doped Sr-214.  Data was collected at $5$ K with the magnetic field oriented within the $ab$-plane.  Solid symbols denote data acquired during sweeping field from $-6$ T to $6$ T (only $-2$ T to $2$ T shown for clarity) while open symbols denote data acquired sweeping field from $6$ T to $-6$ T.  Inset shows subtracted (negative field ramp minus positive field ramp) data for the $x=0.02$ sample. (b) Inverse of the spin susceptibility, minus a constant Pauli term, plotted as a function of temperature for the $x=0$ and $x=0.06$ samples.  Solid lines are Curie-Weiss fits to high temperature data as described in the text.}
\end{figure}

Curie-Weiss fits of the form $\chi(T)=\frac{C}{T-\Theta}+\chi_0$ plotted in Fig. 6 (b) support the notion that magnetic order is suppressed via electron doping, yet local moments ultimately survive at the highest doping levels.  Effective Ir local moment values were determined from the constant $C=\frac{\mu_{eff}^2}{3k_B}$ and evolve from $\mu_{eff}=0.53\pm0.05$ $\mu_B$ in the parent system (consistent with earlier reports \cite{PhysRevB.57.R11039}) to $\mu_{eff}=0.57\pm0.03$ $\mu_B$ in $x=0.06$, while the CW temperature is suppressed from $\Theta=241$ K to $\Theta=105$K as the ferromagnetism associated with the canted AF state is weakened.  We note here however that this is only an approximate metric for quantifying the magnetism of this system; much larger AF exchange couplings and short-range correlations are known to persist above $T_{AF}$ for the parent system and likely also survive above $T_{AF}$ for lightly La-doped concentrations.  The CW fits should therefore be regarded only as a relative gauge for how the strength of magnetic interactions evolve upon carrier substitution. 

As a further probe of the freezing transition at $T_F$, low temperature AC susceptibility measurements were performed on an $x=0.01$ sample, where DC measurements show $T_F=13$ K.  Fig. 7 (a) shows a comparison of the low temperature ZFC cusp in the real part of the susceptibility $\chi^{\prime}$ for 10 and 1000 Hz overlaid with DC magnetization data collected under ZFC. Fig. 7 (b) shows the accompanying frequency dependent peak in the dissipative $\chi^{\prime\prime}$ channel as the system is cooled through $T_F$. The peak of the ZFC cusp shifts to higher temperatures with increasing probe frequency, consistent with the formation of a low temperature spin glass state at $T_F$.  The development of this second, dynamic, component within the low temperature magnetization of electron-doped Sr-214 likely arises from the partial fragmenting of the parent state's AF order into small glassy clusters that survive until at least $x=0.06$.   The magnitude of the frequency shift gives an unusually large value $K=\frac{1}{T_F}\frac{\Delta T_F}{\Delta log(f)}\approx 0.07$ relative to canonical spin glasses \cite{mydosh}, suggesting the dynamics are driven by interacting CAF clusters fragmented from the parent SOM state. 

\section{Magnetotransport Measurements}
In order to search for coupling of the $T_F$ transition to charge transport within La-doped Sr-214, transverse magnetoresistance (MR) measurements were performed with current driven within the $ab$-plane.   MR data collected on a series of parent and La-doped samples are plotted as a percentage $MR=100\times\frac{\rho(H)-\rho(0)}{\rho(0)}$ in Fig. 8. Looking first at undoped Sr-214, low temperature MR data with $H\parallel ab$ reflect the metamagnetic transition at $H=H_C$ consistent with earlier reports \cite{PhysRevB.84.100402}.  Once La is doped into the system, this abrupt decrease in MR is broadened into a nonsaturating, negative linear response at high field. Transverse MR measurements collected with $H\parallel c$ yielded a substantially reduced, negative MR as shown in Fig. 8 (a).  This weak residual negative MR response in the presence of an out-of-plane field may arise from a variety of effects such as weak localization or modified hopping conduction; however the dominant MR with $H\parallel ab$ is driven by charge transport coupling to magnetic order/spin fluctuations within the system.     

\begin{figure}
\includegraphics[scale=.3]{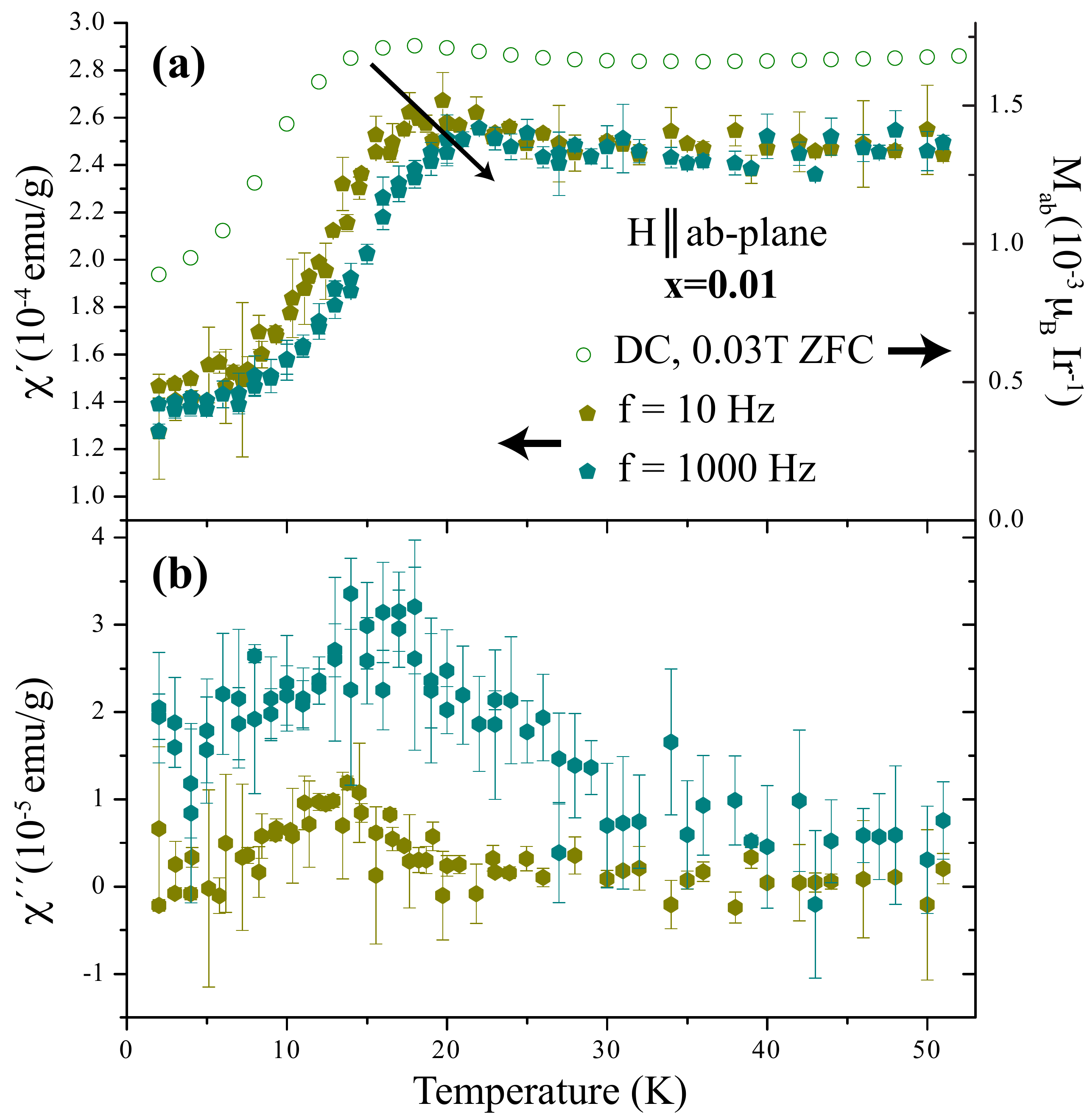}
\caption{Susceptibility data showing the frequency dependence of $T_F$ in $x=0.01$ La-doped Sr-214. (a) Overplot of DC magnetization data collected under ZFC (open circles) and AC $\chi^{\prime}$ collected at $f=10$ and $1000$ Hz (closed pentagons).  (b)  AC susceptibility $\chi^{\prime\prime}$ data collected at $f=10$ and $1000$ Hz.}
\end{figure}

\begin{figure}
\includegraphics[scale=.425]{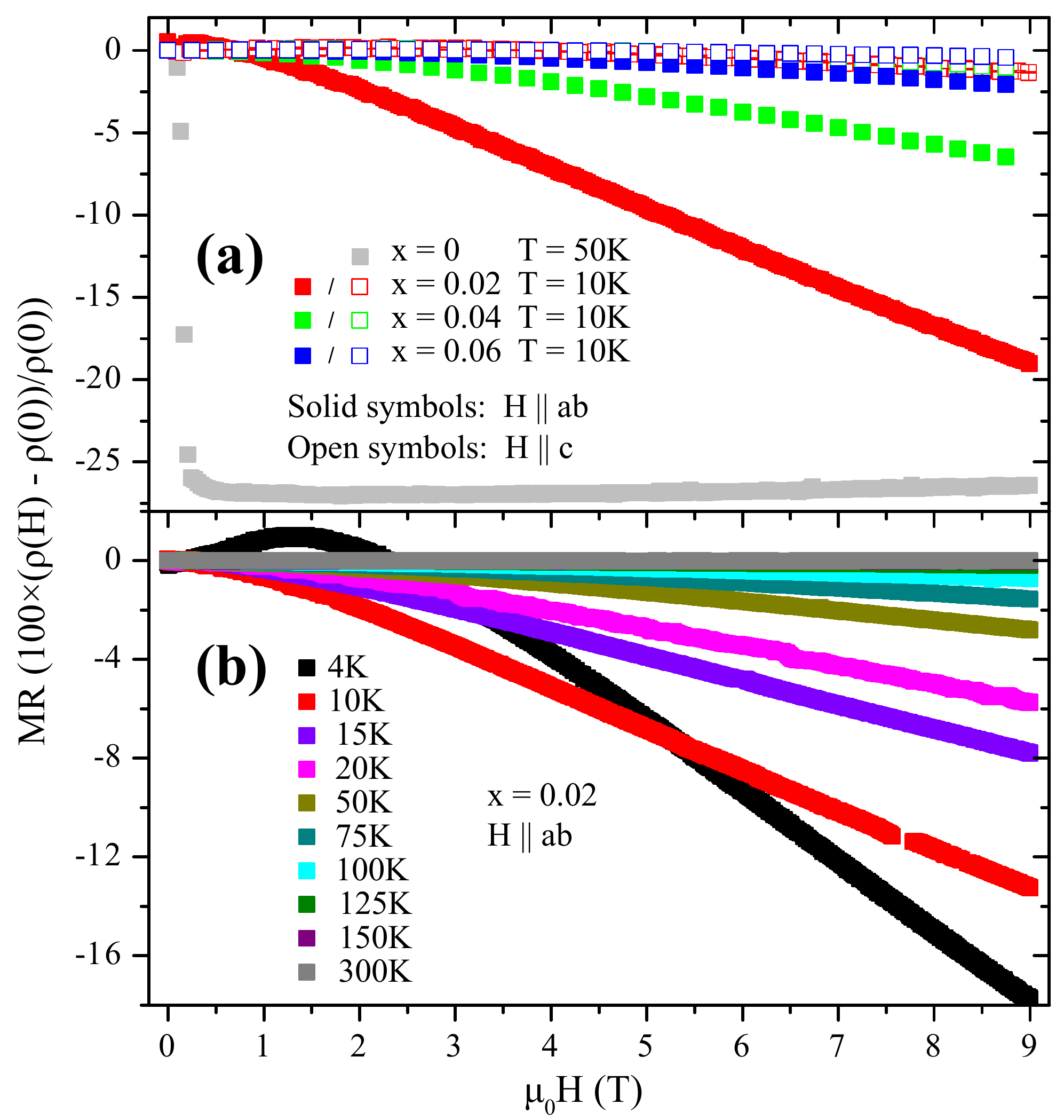}
\caption{Transverse magnetoresistance measurements collected as a function of field at various temperatures for La-doped Sr-214.  (a) MR data collected at temperatures just above the $T_{F}$ state with the field oriented both parallel to the $ab$-plane (solid symbols) and parallel to the c-axis (open symbols). (b) Field dependent MR data collected at various temperatures for the $x=0.02$ concentration with $H\parallel ab$-plane.}
\end{figure}

\begin{figure}
\includegraphics[scale=.370]{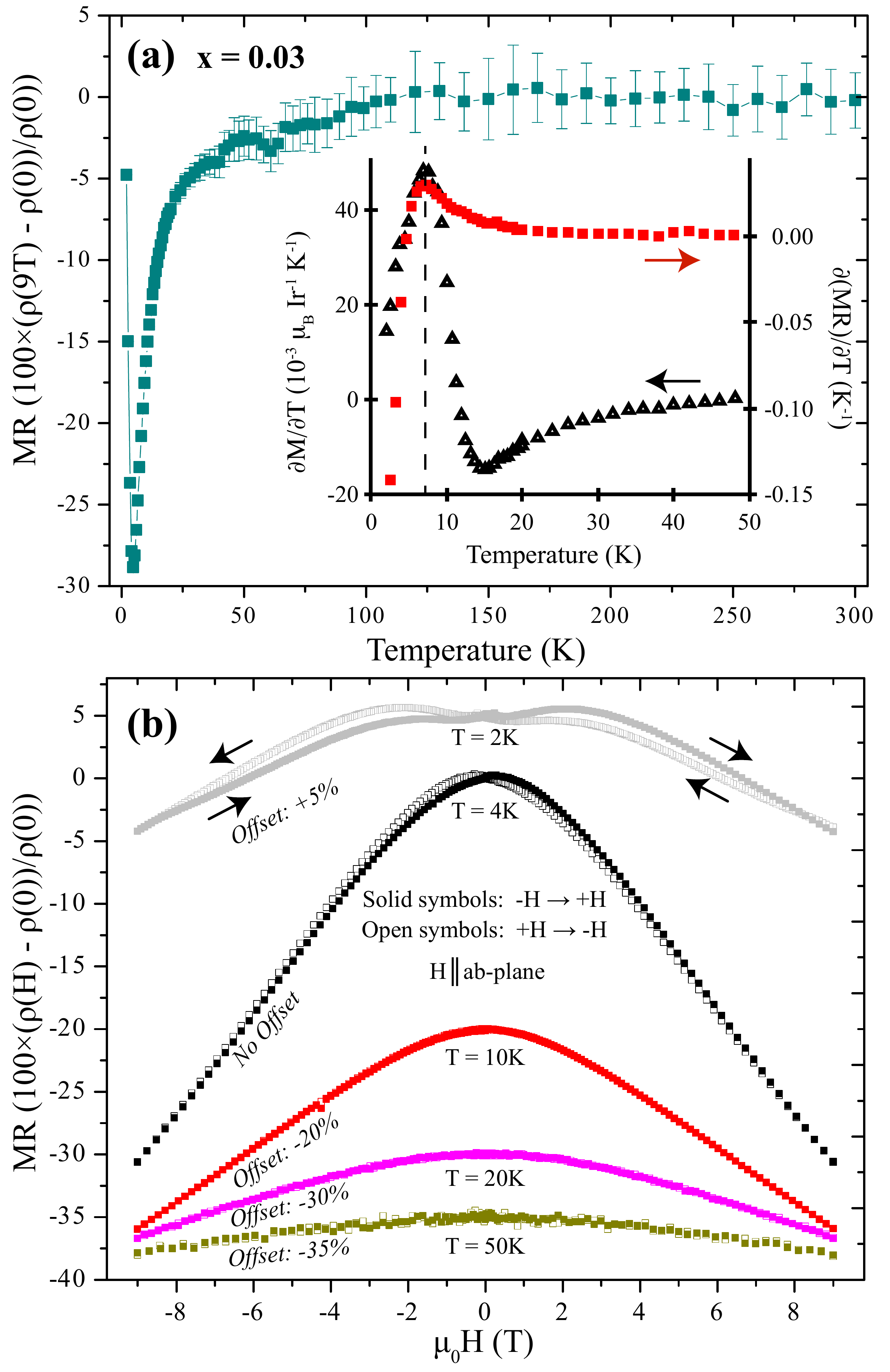}
\caption{Transverse magnetoresistance measurements collected at various fields and temperatures for $x=0.03$ La-doped Sr-214.  (a) Temperature dependence of the high field MR with the $H\parallel ab$-plane. Inset shows overplot of $\frac{\partial MR}{\partial T}$ with the derivative of magnetization $\frac{\partial M}{\partial T}$ on this same sample as described in text.   (b) Field loops of MR data collected at various temperatures with $H\parallel ab$-plane.  Data is offset for clarity.}
\end{figure}

In doped samples at low values of $H_{ab}$, a remnant negative cusp in MR develops at $H\approx H_C$ for $T_F<T<T_{AF}$.  Upon cooling below $T_{F}$, this negative cusp is replaced with a positive upturn in the low field MR as shown by the temperature dependence of MR response for $x=0.02$ plotted in Fig. 8 (b).  This low field upturn is consistent with the onset of a spin glass phase below $T_{F}$ where the in-plane field initially "unlocks" moments from the frozen state \cite{AuFe,NiMn,AgMn} and enhances spin-flip scattering in transport.  Upon continued increase in field, the negative MR driven by polarizing domains/spin fluctuations associated with the canted AF order in the sample dominates---leading to a complex interplay between weak ferromagnetism and this low temperature glass state.   Similar behavior has also been observed in nanoscale phase segregated colbaltites where spin dependent transport between ferromagnetic clusters results in a maximum in MR at the coercive field  \cite{wu2005intergranular}. 

This is further illustrated in MR data collected on an $x=0.03$ sample ($T_F=11$ K) plotted in Fig. 9.  $\rho(T)$ data measured during warming in both $0$ T and $9$ T in-plane oriented fields were used to determine the temperature dependence of the MR in Fig. 9 (a).  The magnitude of the negative MR ratio grows gradually upon cooling and eventually diverges as it approaches $T\approx 7$ K.  Further cooling results in the negative MR response being rapidly suppressed as moments in the sample freeze.  The inset in Fig. 9 (a) overplots the temperature derivative of this MR response ($\frac{\partial MR}{\partial T}$) with that of the low-field magnetization ($\frac{\partial M}{\partial T}$) for this same sample.   The coincident peak in the thermal derivatives of MR and M(T) correlates the two phase behaviors; however some care should be used in directly comparing onset temperatures via the two probes.  In particular, fields required to generate an appreciable MR response also suppress the onset temperature $T_{F}$, and the high-field MR response (plotted in Fig. 9 (a)) may only reflect relative changes between a highly polarized $T_{F}$ state and the non saturating negative linear response above $T_{F}$.      

A transition into a frozen state below $T_{F}$ is further illustrated by hysteretic behavior in MR data.  Fig. 9 (b) illustrates this hysteresis in MR by plotting field loops in the same $x=0.03$ sample at select temperatures both above and below $T_{F}$. As the system is cooled below $T_{F}$, the MR with $H\parallel ab$ switches from one reflective of a nearly reversible polarization of canted AF moments to an irreversible switching of a frozen state.   The onset of hysteresis is accompanied by the development of a low field positive MR response which peaks around $2.5$ T at $2$ K before reverting to a negative linear response at higher field values.  The field value where the MR peaks is temperature dependent and shifts toward $0$ T as $T_{F}$ is approached.  This is broadly consistent with the development of a larger volume fraction of glassy spins and stronger coupling between frozen moments as the system is cooled deeper into a spin glass transition.    

\section{Neutron Scattering Measurements}
In order to explore the evolution of long-range AF order upon La-substitution, neutron diffraction measurements were performed.  Neutron data are summarized in Fig. 10.  Looking first at Fig. 10 (a), momentum scans show the rapid suppression of the strongest magnetic Bragg peak at \textbf{Q}$=(1, 0, 2)$ as electrons are doped into the system.  Here the background subtracted intensities of the (1, 0, 2) peaks normalized by (0, 0, 4) nuclear Bragg reflections are plotted for both $x=0.01$ and $x=0.02$ samples.  Magnetic scattering along the parent system's ordering wave vector is barely discerned in the $x=0.02$ sample, and the weak residual peak was confirmed to be magnetic via its temperature dependence plotted in Fig. 10 (b).  A sample with $x=0.04$ was also measured; however no magnetic Bragg peak could be resolved at the (1, 0, 2) or equivalent positions.  This demonstrates the complete suppression of the long-range canted AF state endemic to the parent Sr-214 system within the detection limits of the measurement ($\approx 0.06$  $\mu_B$).  

Our neutron measurements combined with the survival of irreversibility in the static spin susceptibility measurements of all samples implies a transition into a short-range ordered state for $x> 0.03$.  Diffuse scattering from such a state would be below the detection limit of the present measurements.  Fig. 10 (c) summarizes the evolution of the ordered AF moment obtained via our neutron data.  For comparison, ordered moment estimates were also generated via the high field (6T) magnetization data (Fig. 6) assuming canted antiferromagnetic moments polarized across the metamagnetic phase boundary and spin canting angles locked to in-plane octahedral rotation angles (refined for each concentration).  For samples where long-range order remains resolvable in diffraction, this bulk magnetization derived estimate for ordered AF moments agrees well with the neutron data.  For $x>0.03$ the remaining weak ferromagnetism in magnetization data then allows estimates of AF moments associated with remnant short-range order not apparent in the neutron data.  

Using this moment estimate derived from magnetization measurements, data for the $x=0.04$ concentration suggest that long-range AF order should be within the resolution of the neutron measurements. The absence of observable AF order in scattering measurements from this same sample therefore demonstrate a transition to a short-range ordered state at this concentration---one where the momentum broadening pushes magnetic diffraction intensities below the detection limit. Extending this analysis to higher La-concentrations, the survival of weak ferromagnetism in magnetization data associated with short-range AF order at $x=0.06$ suggests a scenario where AF order survives in local pockets within an electronically phase separated ground state up to the doping limit of La within Sr-214. 

\begin{figure}
\includegraphics[scale=.18]{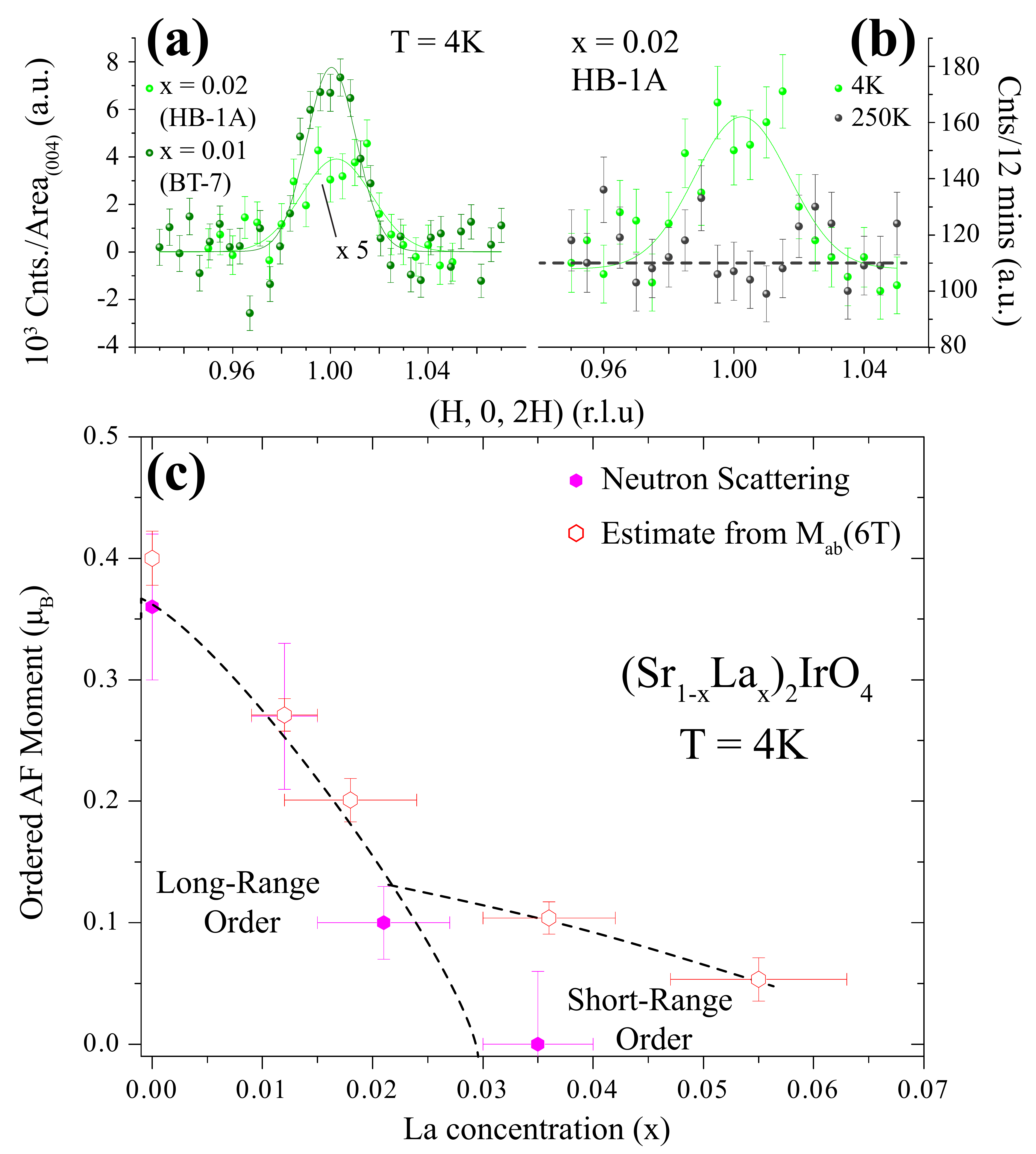}
\caption{Neutron scattering data showing momentum scans through the magnetic Bragg position \textbf{Q}$=(1, 0, 2)$. (a) Radial Q-scans through the AF Bragg peak at (1, 0, 2) collected at 4 K for both $x=0.01$ and $x=0.02$.  Data has nonmagnetic background contribution removed, been normalized by the radially integrated area of the (0 0 4) nuclear Bragg reflection. Intensity of data for $x=0.02$ has been multiplied by 5 for clarity.  (b) Radial scans through the (1, 0, 2) position for the $x=0.02$ sample at both 4 K and 250 K. (c) Evolution of the low temperature ordered AF moments in La-doped Sr-214.  Closed symbols denote the ordered AF moment measured directly with neutron scattering data and open symbols denote an inferred ordered moment assuming the canted spin structure of Sr-214, an approximately saturated canted magnetization at 6T, and the refined octahedral canting angles for each concentration.  Dashed lines mark the suggested boundaries between short and long-range AF order.}
\end{figure}

\section{Scanning Tunneling Microscopy Measurements}
Low-temperature scanning tunneling microscopy and spectroscopy (STM/S) measurements were performed to explore whether electronic phase separation occurs in La-doped Sr-214.  A typical STM topography for an $x=0.05$ sample is shown in Fig. 11 (a). The crystal cleaves between adjacent SrO planes, and from previous experiments \cite{Okada} we expect that the maxima in the STM topographies are associated with the Sr atoms. Correspondingly, in Fig. 11 (a) the La dopants appear as dark-centered bright squares \cite{hogan}, whose number is consistent with the $x=0.05$ nominal doping of the sample. 

Insulating patches (e.g., at the upper right corner of Fig. 11 (a)) are visible as dark depressions in the topography, where the low local density of states forces the tip to push in so as to maintain a constant tunneling current. To examine the local density of states more closely, $\frac{\partial I}{\partial V}$ spectra were collected over on a dense grid ($0.9$ spectra/$\AA^2$) in a $15\times15$ nm$^2$ area with an energy range $-0.6$ to $+0.6$ eV. The results are summarized in Figs. 11 (b) and (d). Spectra far from La dopants show a large, $\approx 300$ mV insulating gap, but clusters of La dopants show metallic behavior, with finite density of states at $E_F$. The evolution of the $\frac{\partial I}{\partial V}$ spectra between these two extremes is shown in Fig. 11 (d). 

To quantify the inhomogeneity in the density of states we produced a ``gap map" shown in Fig. 11 (b), where the gap is defined as the energy range in each spectrum in which the measured $\frac{\partial I}{\partial V}$ was zero within a small noise threshold.  In Fig. 11 (c) we show the ``local La doping" in the same area as the $\frac{\partial I}{\partial V}$ map, obtained from counting the La dopants in the STM topography.  Here each dopant is represented as a normalized Gaussian with $\sigma = 1.5$ nm. The metallic spectra are somewhat correlated with the La dopant density, with a correlation coefficient between Figs. 11 (b) and (c) of $R \approx 0.4$. The observed electronic structure is potentially also influenced by La dopants in the second SrO plane from the surface, which are not seen with STM.

\begin{figure}
\includegraphics[scale=.38]{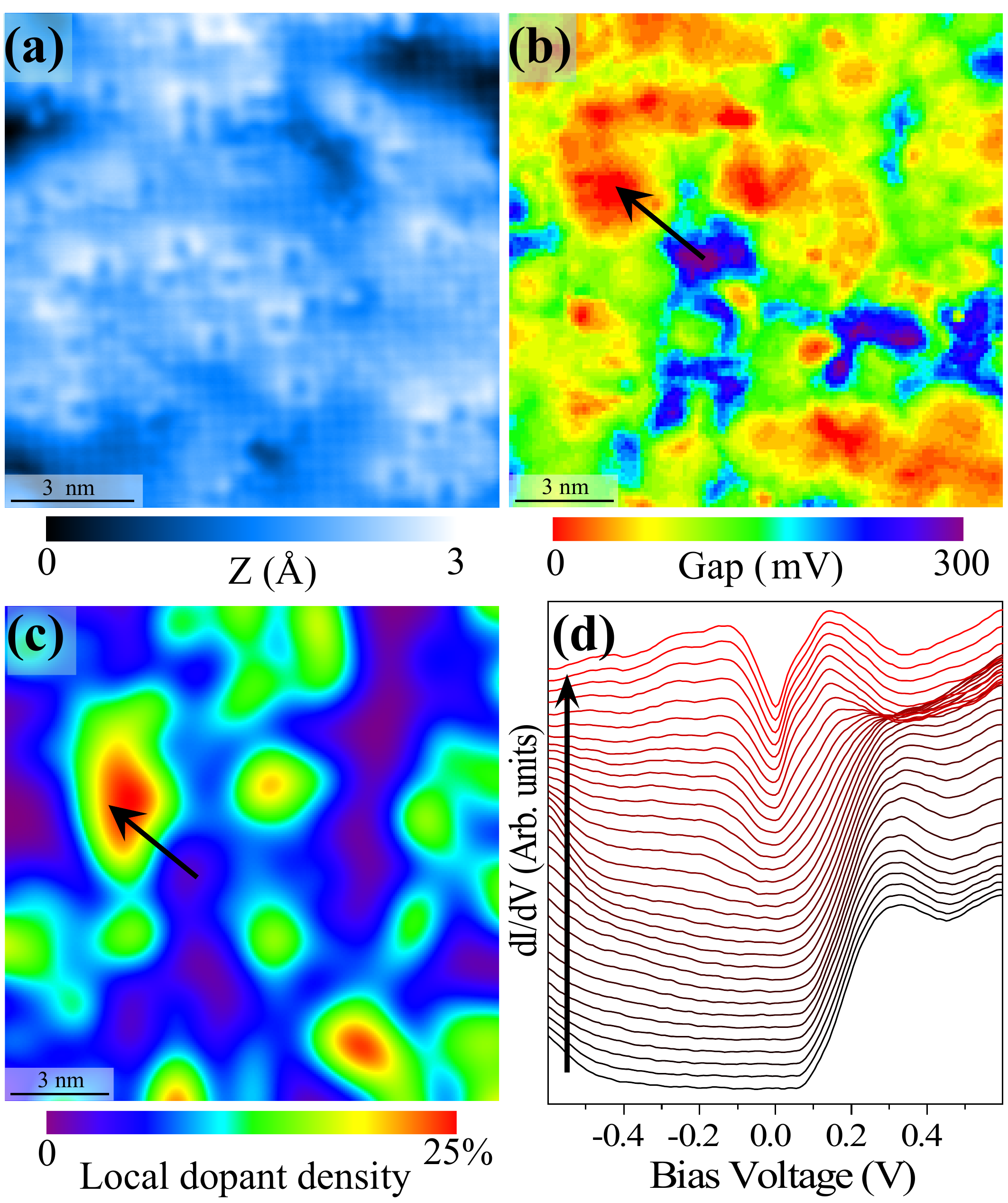}
\caption{(a) A representative, $12 \times 12$ nm$^2$ STM topography of a sample with $x\approx 0.05$ nominal La doping, taken at $V_b$ = +300 mV at $I_{set} = 80$ pA. (b) Gap map extracted from a $\frac{\delta I}{\delta V}$ map, $15 \times 15$ nm$^2$, taken on the same sample as in (a) but over a different area. (c) Dopant density map taken in the same area as (b); dopants were counted visually. (d) Line cut showing the evolution of the $\frac{\delta I}{\delta V}$ spectra from fully gapped (purple) to no gap (red). The position and orientation of the line cut is indicated by the black arrows in (b) and (c).}
\end{figure}

\section{Discussion}
The global picture provided by a combined analysis of the scattering, transport, STM, and magnetization data demonstrates that electron doping into the SOM state of Sr-214 renders an electronically phase separated ground state.  This separation into nanoscale patches of doped charge and persistent insulating regions persists beyond the disappearance of long-range AF order and argues for a percolative origin to the sharply reduced resistivity of Sr-214 upon alloying La into the lattice.  For La-doped Sr-214 with $x\geq 0.04$, this percolation mechanism leads to a high temperature metallic state whose conduction switches to weakly insulating at low temperatures.  

The definitive origin of this persistent low temperature upturn remains elusive.  Traditional models of weak localization driven transport fail to fully capture the resistivity data in this regime \cite{PhysRevLett.42.673, RevModPhys.54.437}.  Additionally, the inflection in $\rho(T)$ marking the transition from weakly insulating to metallic behavior (ie. where $\frac{\partial  \rho}{\partial T}$ changes sign) of the highest measured La concentration ($x=0.06$) at $T\approx 100$K does not directly coincide with the onset of irreversibility in the low field susceptibility at $T_{AF}\approx 125$ K.  We note here that this minimum in conductivity decoupled from the onset of AF order is similar to the tunneling transport behavior observed in other lightly doped Mott states such as the high-T$_C$ cuprates \cite{PhysRevLett.87.017001,RevModPhys.70.897}.  A recent angle-resolved photoemission study's report of a pseudogap in La-doped Sr-214 is also consistent with our observation of a gradual gapping out of conductivity at low temperatures \cite{2015arXiv150600616D}.   In analyzing the metallic behavior at high temperatures within the two-dimensional limit of planar transport, the high temperature resistivity for the $x=0.06$ sample appears to violate the Mott limit for metallic transport where $k_fl\approx0.6<1$ \cite{Mott}.  This is likely reflective of the phase segregated nature of the electronic ground state where only a fraction of the sample volume contributes to conduction.

The survival of nanoscale insulating regions at $x=0.05$ (near the limit of La-substitution) as seen via STM measurements demonstrates that the doping driven enhancement in conductivity in charge transport data originates from percolative conduction channels.  The growth of regions with enhanced conductivity rapidly suppresses long-range AF order in the system beyond $x\approx0.03$, although short-range AF order seemingly survives up to the doping limit of La within Sr-214.  More broadly, the survival of static AF order combined with gapped regions in STM spectra demonstrates that $x=0.06$ La substitution is insufficient to destabilize the parent SOM state and drive a global MIT.  This observation is consistent with a recent ARPES study showing a local distribution of gap values accounting for the in-gap spectral weight of La-doped Sr-214 \cite{2015arXiv150308120B}. 

Interestingly, upon electron doping a low temperature state spin glass-like order appears in both bulk magnetization and magnetotransport measurements below an energy scale $T_{F}$.  While this state is most evident under light electron substitution, signatures associated with this frozen state also appear to some degree in undoped Sr-214 samples.  This incipient spin glass behavior likely arises from defect states and impurity doping within the nominally undoped parent material. Crystals of parent Sr-214 are prone to low levels of impurity doping via defects and slight off stoichiometry in oxygen content.  These can be healed somewhat via post growth annealing; however they are difficult to remove entirely.  The broadened $T_{F}$ transition in parent Sr-214 as well as the apparent enhancement in the onset temperature (Fig. 4) are consistent with the modification of a spin glass state in the presence of disorder.  This supports an impurity driven origin of this freezing behavior in the parent material.     

A magnetic phase diagram summarizing our measurements is plotted in Fig. 12.   Onset temperatures for AF order obtained via bulk magnetization and neutron scattering measurements are plotted with a dashed line denoting the crossover from long-range to short-range AF order.  The onset temperatures $T_{F}$ are also plotted as a low temperature phase boundary. Given the recent parallels drawn between electron-doped Sr$_2$IrO$_4$ and hole-doped cuprates \cite{PhysRevLett.106.136402}, it is worthwhile to consider parallels between the electronic phase diagrams of the two systems. In particular, Sr-214's iridate phase diagram is somewhat reminiscent of the hole-doped phase diagram of its 3d-analog La$_2$CuO$_4$ \cite{RevModPhys.70.897}.   In both systems, light carrier substitution leads to a phase separated state where AF is rapidly suppressed.  As the parent system's AF order is degraded, an intermediate spin glass state appears in both phase diagrams prior to the onset of a global metallic state.  It is worth noting however that the suppression of $T_F$ with continued doping is much more gradual in electron-doped Sr-214 than in hole-doped La$_2$CuO$_4$. 

\begin{figure}
\includegraphics[scale=.375]{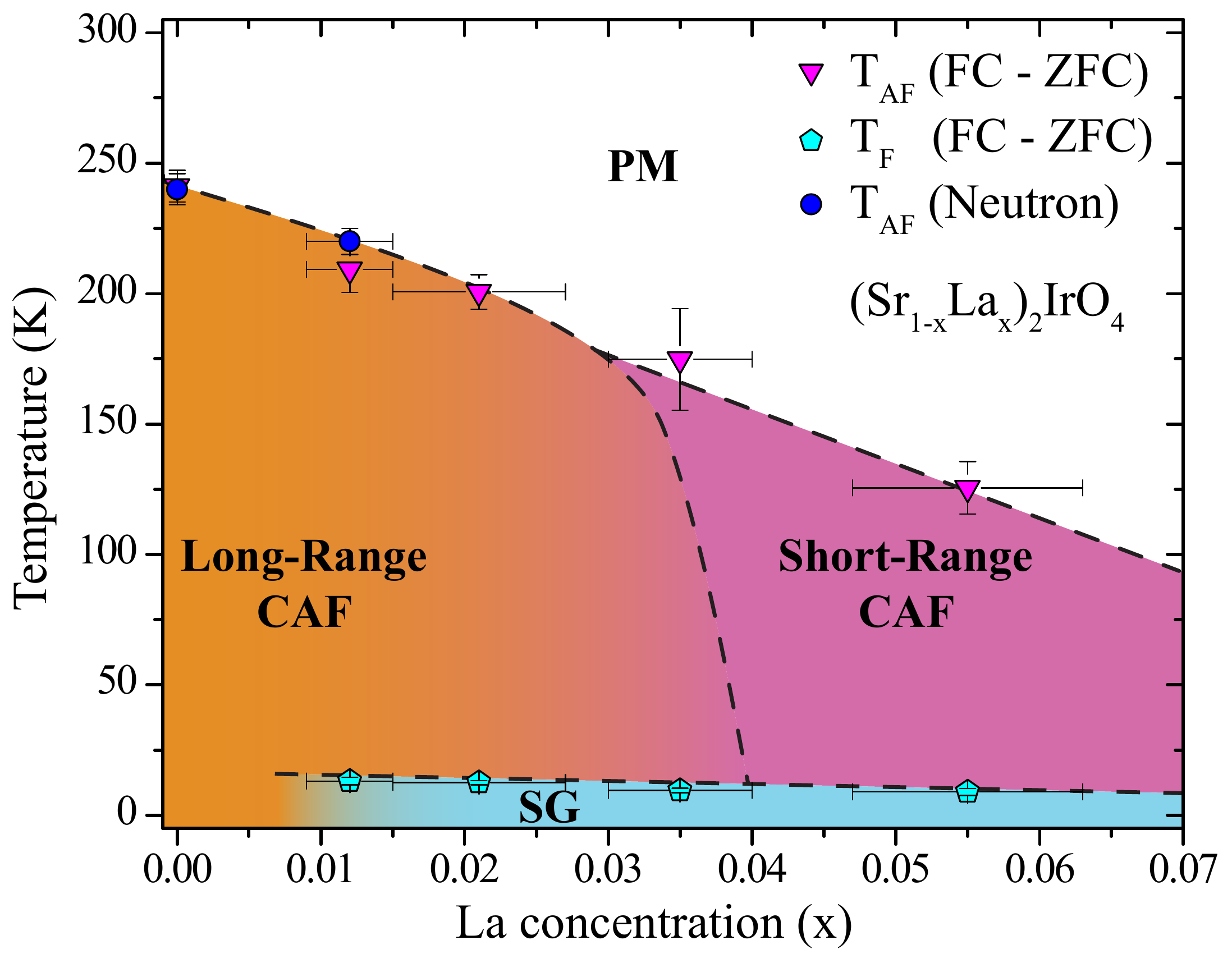}
\caption{Evolution of magnetic order in La-doped Sr-214.  Blue circles denote the AF ordering transition measured directly with neutron scattering data, and pink triangles denote the onset of irreversibility in magnetization data.  Dashed lines mark the suggested boundaries between short and long-range AF order.  The proposed low temperature spin glass phase boundary is denoted by cyan pentagons. }
\end{figure}

Unlike the hole-doped cuprates, short-range AF order persists over a wide doping range in La-doped Sr-214.   This also contrasts the electron-doped phase diagram of (Sr$_{1-x}$La$_{x}$)$_3$Ir$_2$O$_7$ where static AF order abruptly vanishes across the MIT phase boundary \cite{hogan}.  Within both La-substituted Sr-214 and Sr-327, light electron doping renders an electronically phase separated ground state with coexisting metallic and gapped regions; however, beyond a threshold value of $6\%$ electrons per Ir in Sr-327, AF order collapses into a globally gapless metal.  In Sr-214, continued electron doping beyond $6\%$ electrons per Ir fails to fully destabilize the SOM state, and a phase separated state persists until the limit of La-substitution ($\approx 12\%$ electrons per Ir). This is likely due to either a nearby competing electronic instability in Sr-327 \cite{hogan} or simply due to the inherently stronger Mott state of Sr-214.  

\section{Conclusions}
Combined neutron scattering, magnetotransport, magnetization, and STM/S data reveal a complex evolution of the electronic properties of the spin-orbit Mott ground state of (Sr$_{1-x}$La$_x$)$_2$IrO$_4$.  Electron-doping results in  electronic phase separation into coexisting metallic and insulating nanoscale regions and percolative transport. Long-range AF order collapses beyond $x\approx 0.03$.  Short-range AF order however survives and persists up to the upper limit of La-substitution in an electronically phase segregated ground state.  Furthermore, electron doping stabilizes a new low temperature phase transition consistent with the formation of a spin glass-like state likely comprised of frozen clusters of CAF nanoscale domains.  This spin glass phase is necessarily intermediate to the formation of a globally gapless metallic ground state in electron-doped Sr-214, reminiscent of the electronic phase diagram of monolayer hole-doped high-$T_C$ cuprates.  

\acknowledgments{
X.C. gratefully acknowledges Brandon Isaac and the 11-BM beam line staff for assistance with experiments.  This work was supported in part by NSF CAREER award DMR-1056625 (S.D.W. and X.C.). This work utilized facilities supported in part under NSF award DMR-0944772 and SQUID measurements were supported in part by grant DMR-1337567.  Part of this effort was supported by the US Department of Energy (DOE), Office of Basic Energy Sciences (BES), Materials Sciences and Engineering Division, (TZW).  STM work was supported by NSF DMR-1305647 (V.M.).  Use of the Advanced Photon Source at Argonne National Laboratory was supported by the U. S. Department of Energy, Office of Science, Office of Basic Energy Sciences, under Contract No. DE-AC02-06CH11357.  Research conducted at ORNLs High Flux Isotope Reactor was sponsored by the Scientific User Facilities Division, Office of Basic Energy Sciences, US Department of Energy.  The identification of any commercial product or trade name does not imply endorsement or recommendation by the National Institute of Standards and Technology.  }

\bibliography{LaDopedBib}

\end{document}